\documentclass[12pt]{iopart}
\usepackage{subfigure}
\usepackage{graphicx}
\usepackage{dcolumn}
\usepackage{graphics}
\usepackage{color}
\usepackage{bm}
\usepackage{epsfig}

\begin{document}

\rapid{On the magnetic structure of $\mathrm{Sr_{3}Ir_{2}O_{7}}$: an x-ray resonant scattering study}
\author{S Boseggia$ˆ1,2$, R Springell$ˆ3$, H C Walker$ˆ4$, A T Boothroyd$ˆ5$, D Prabhakaran$ˆ5$, S P Collins$ˆ2$ and D F McMorrow$ˆ1$}
\address{$ˆ1$ London Centre for Nanotechnology and Department of Physics and Astronomy, University College London, London WC1E 6BT, United Kingdom}
\address{$ˆ2$ Diamond Light Source Ltd, Diamond House, Harwell Science and Innovation Campus, Didcot, Oxfordshire OX11 0DE, UK}
\address{$ˆ3$ Royal Commission for the Exhibition of 1851 Research Fellow, Interface Analysis Centre, University of Bristol, Bristol BS2 8BS, United Kingdom}
\address{$ˆ4$ Deutsches Elektronen-Synchrotron DESY, 22607 Hamburg, Germany}
\address{$ˆ5$ Clarendon Laboratory, Department of Physics, University of Oxford, Parks Road, Oxford OX1 3PU, United Kingdom}
\ead{stefano.boseggia@diamond.ac.uk}

\begin{abstract}
This report presents azimuthal dependent and polarisation dependent x-ray resonant magnetic scattering at the Ir L$_{3}$ edge for the bilayered iridate compound, Sr$_{3}$Ir$_{2}$O$_{7}$. Two magnetic wave vectors, $\textbf{k}_{1}$=($\frac{1}{2}$,$\frac{1}{2},$0) and $\textbf{k}_{2}$=($\frac{1}{2}$,-$\frac{1}{2},$0), result in domains of two symmetry-related G-type antiferromagnetic structures, noted A and B, respectively. These domains are approximately 0.02$\,$mm$^{2}$ and are independent of the thermal history. An understanding of this key aspect of the magnetism is necessary for an overall picture of the magnetic behaviour in this compound. Azimuthal and polarisation dependence of magnetic reflections, relating to both magnetic wave vectors, show that the Ir magnetic moments in the bilayer compound are oriented along the \textit{c} axis. This contrasts with single layer Sr$_{2}$IrO$_{4}$ where the moments are confined to the \textit{ab} plane.
\end{abstract}
\submitto{\JPCM}
\maketitle

\date{\today}

The 5d transition metal oxides are a new frontier for research into compounds where spin-orbit coupling and crystal field effects compete on an equal footing. A host of particularly interesting physical phenomena may occur at the crossroads of these two energy scales \cite{Kim2,Shitade,Jiang,Machida}. For the case of Ir$^{4+}$ ions, where relativistic effects are significant and for a low-spin d$^{5}$ configuration in the presence of an octahedral crystal field, a four-fold degenerate $J_{eff}=3/2$ multiplet and a two-fold degenerate $J_{eff}=1/2$ multiplet are created. The lower lying $J_{eff}=3/2$ quartet is filled and the single remaining electron then occupies the $J_{eff}=1/2$ band, which can then be split into an upper and a lower Hubbard band by 5d electron correlations, forming an insulating gap. The n=2 member of the Sr$_{n+1}$Ir$_{n}$O$_{3n+1}$ Ruddlesden-Popper series, Sr$_{3}$Ir$_{2}$O$_{7}$, is a particularly interesting Ir$^{4+}$ compound; it is only weakly insulating and its proximity to a metal insulator transition might, at first thought, be enough to perturb the $J_{eff}=1/2$ states. A recent x-ray resonant scattering study has shown this not to be the case \cite{Boseggia} and it bears many of the same hallmarks as its single layered counterpart, Sr$_{2}$IrO$_{4}$ \cite{Kim2}. In fact, similar to Sr$_{2}$IrO$_{4}$, Sr$_{3}$Ir$_{2}$O$_{7}$ is a G-type antiferromagnet at low temperature \cite{Boseggia,Kim3,Dhital}. However, there remains some controversy over the details of the precise magnetic structure reported in the recent literature, and explanations of certain aspects of the bilayered system that are present in the experimental data \cite{Boseggia, Kim3, Dhital, Jackeli}. This article reports an investigation of the antiferromagnetic structure with an in-depth study of the azimuthal and polarisation dependent resonant scattering signal with particular consideration of the domain behaviour, resulting as a consequence of the degeneracy of two magnetic propagation vectors $\textbf{k}_{1}$=($\frac{1}{2}$,$\frac{1}{2},$0) and $\textbf{k}_{2}$=($\frac{1}{2}$,-$\frac{1}{2},$0) \cite{Boseggia}.

\begin{figure}[h!]
\centering
\includegraphics[width=0.75\textwidth,bb=120 365 460 510,clip]{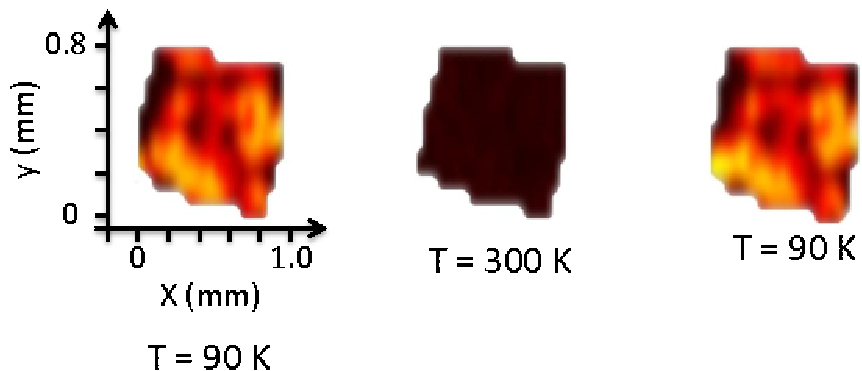}\caption
{\label{figure1}(Color online) The intensity of the ($\frac{1}{2}$,$\frac{1}{2}$,24) reflection, corresponding to the $\textbf{k}_{1}$=($\frac{1}{2}$,$\frac{1}{2},$0) magnetic propagation vector, has been measured as a function of x and y sample position in a temperature sequence T=90$\,$K, 300$\,$K, 90$\,$K. The bright yellow regions are areas of high intensity and the dark areas are close to zero intensity.}
\end{figure}

Sr$_{3}$Ir$_{2}$O$_{7}$ is a weakly insulating magnetic compound which crystallises in the tetragonal I4/mmm crystal structure with $\mathrm{a=3.897(5)\,\AA}$ and $\mathrm{c=20.892(5)\,\AA}$ at room temperature \cite{Subramanian}. Others have reported more complicated, orthorhombic structures \cite{Cao,Matsuhata}, which describe a correlated rotation of IrO$_{6}$ octahedra, but we could not detect any evidence for this in our sample \cite{Boseggia} and the tetragonal representation is a sufficient starting point from which to investigate the magnetic structure. Bulk magnetisation data in the 5$\,$K to 300$\,$K temperature interval reveals three distinct transitions at $\sim$280$\,$K, $\sim$230$\,$K and $\sim$50$\,$K \cite{Cao,Nagai,Boseggia}, although it should be noted that the precise values of these transitions vary by as much as 10$\,$K from study to study. Our previous x-ray scattering investigation of this compound \cite{Boseggia}, combined with calculations performed using the SARA\textit{h} program \cite{Wills}, which calculates irreducible representations based on symmetry analysis and group theory, indicated that Sr$_{3}$Ir$_{2}$O$_{7}$ displays a two-domain G-type antiferromagnetic structure with wave vectors $\textbf{k}_{1}$=($\frac{1}{2}$,$\frac{1}{2},$0) and $\textbf{k}_{2}$=($\frac{1}{2}$,-$\frac{1}{2},$0). These two modulation vectors are represented by symmetry-related structures, denoted A and B in Figure 6 of Boseggia \textit{et al.} \cite{Boseggia}, and result in magnetic Bragg peaks at ($\frac{1}{2}$,$\frac{1}{2}$,L=even) and ($\frac{1}{2}$,$\frac{1}{2}$,L=odd) positions, respectively. By using a highly focussed x-ray beam and rastering over the surface of the sample, we were able to `image' domains, relating to these two wave vectors. Figure \ref{figure1} shows the domain pattern for the ($\frac{1}{2}$,$\frac{1}{2}$,24) peak, rastered over an area of approximately 1$\,$mm$^{2}$ at 90$\,$K, well below the transition temperature, heated above the N\'{e}el transition to 300$\,$K and then cooled back to 90$\,$K, revealing domains of the order 100$\times$200$\,\mu$m$^{2}$, which are independent of the thermal history. An understanding of this domain behaviour and even the simple fact that there are two possible wave vectors responsible for the magnetic order are essential pre-requisites for an accurate study of the magnetic structure.

Utilising version 2K of the representational analysis program SARA\textit{h} \cite{Wills}, we have exploited the group theory approach to determine the magnetic structures allowed by symmetry. The results of the calculation show 3 possible irreducible representations (IR): $\Gamma_2$, $\Gamma_6$, $\Gamma_8$ that correspond to a G-type antiferromagnet where the moments are oriented along the [001], [-110] and [110] directions, respectively. Using the magnetic structures derived from representational analysis, the azimuthal dependence and the polarisation dependence of the ($\frac{1}{2}$,$\frac{1}{2}$,24) and ($\frac{1}{2}$,$\frac{1}{2}$,23) magnetic reflections have been calculated using the FDMNES computer program \cite{Joly,note}. The following report aims to fully explore the azimuthal and polarisation dependences of magnetic reflections for $\textbf{k}_{1}$=($\frac{1}{2}$,$\frac{1}{2},$0) and $\textbf{k}_{2}$=($\frac{1}{2}$,-$\frac{1}{2},$0) wave vectors, comparing to the calculation discussed above, in order to determine the orientation of the magnetic moments.

\begin{figure}[h!]
\centering
\includegraphics[width=0.5\textwidth,bb=60 230 560 620,clip]{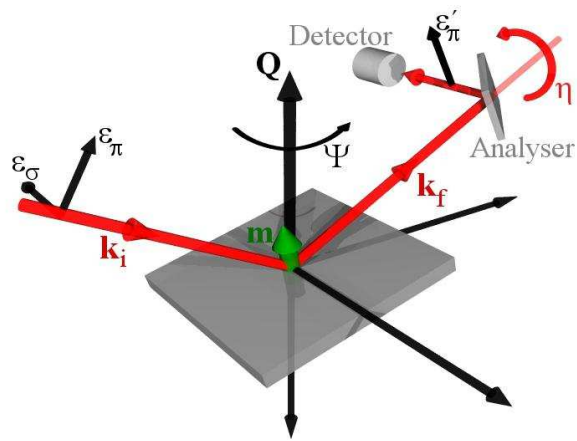}\caption
{\label{figure2}(Color online) The experimental configuration used to perform the azimuthal scans and the polarisation analysis measurements. The sample has been oriented with the (0,0,1) and the (0,1,0) reflections lying in the scattering plane, defined by the incoming and outgoing wave vectors. Azimuthal scans were performed in a vertical scattering geometry ($\sigma$ polarised incident beam) by rotating the sample around the $\Psi$ axis. For domain imaging measurements and the polarisation dependence of the x-ray magnetic scattering, a configuration with the [001] and [010] directions lying in the horizontal scattering plane was used ($\pi$ polarised incident beam). The polarisation of the scattered beam was scanned between $\eta$=0$^{\circ}$ and $\eta$=180$^{\circ}$, where $\eta$=0$^{\circ}$ corresponds to a $\sigma$ polarised beam.}
\end{figure}

The x-ray magnetic scattering experiment was performed at the I16 beamline, Diamond Light Source, Didcot, UK. A monochromatic x-ray beam at the Ir L$_3$ edge (11.217 keV) was obtained by means of a U27 undulator insertion device and a channel-cut Si (111) monochromator. A beam size of 20$\times$200$\,\mu$m (V$\times$H) at the sample position was achieved using a pair of 1.2$\,$m mirrors. A schematic of the experimental set-up is shown in Figure \ref{figure2}.

For the domain imaging measurements and the azimuthal scans the beam size was further reduced to 20$\times$50$\,\mu$m (V$\times$H) by a set of slits.
The sample orientation was controlled by means of a Newport 6-axis N-6050 Kappa diffractometer that allows measurements to be made in both vertical and horizontal
scattering geometries. In order to detect the scattered photons we used two different configurations: a Pilatus 100k detector, and an avalanche photodiode (APD), together with a pyrolytic graphite (008) crystal to analyse the polarisation of the scattered beam. The Pilatus 100k detector has been exploited in the domain imaging measurement (see Figure \ref{figure1}) in order to maximise the data throughput and consequently decrease the acquisition time. For the azimuthal scans and the polarisation analysis measurements the latter configuration has been adopted.

A Sr$_{3}$Ir$_{2}$O$_{7}$ single crystal was mounted on a boron capillary and a nitrogen gas jet cooler was exploited to avoid mechanical vibrations that might produce uncertainty in the sample position with respect to the incident x-ray beam. The sample was oriented with the [001] direction perpendicular to the sample surface and the [010] direction lying in the vertical scattering plane at the azimuthal origin, for the azimuthal measurements ($\sigma$ polarised incident beam). A note of caution should be given at this point, that a large number of multiple scattering peaks were observed during the measurement process, and a great deal of care was taken in order to avoid them.

\begin{figure}[h!]
\centering
\includegraphics[width=0.5\textwidth,bb=45 220 530 600,clip]{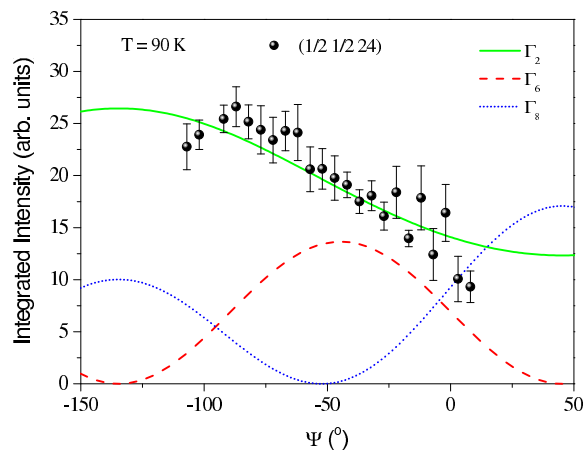}\caption
{\label{figure3}(Color online) Azimuthal dependence of the ($\frac{1}{2}$,$\frac{1}{2}$,24) magnetic reflection at the Ir $L_{3}$ edge in the rotated $\sigma$-$\pi$ polarisation channel at T=90$\,$K.
The black solid points represent the integrated intensity of the experimental ($\frac{1}{2}$,$\frac{1}{2}$,24) magnetic reflection measured rotating the sample around the scattering vector $\textbf{Q}$ by an angle $\Psi$. The green solid line, the red dashed line and the blue dotted line show the FDMNES calculations relative to the IR's $\Gamma_{2}$, $\Gamma_{6}$ and $\Gamma_{8}$, respectively.}
\end{figure}

Figure \ref{figure3} shows the azimuthal dependence of the ($\frac{1}{2}$,$\frac{1}{2}$,24) magnetic reflection at T=90$\,$K, using the resonant enhancement at the Ir $L_{3}$ edge. The sample was rotated about the scattering vector $\textbf{Q}$ by an angle $\Psi$ and the intensity in the rotated $\sigma$-$\pi$ channel was measured. The integrated intensity of this reflection as a function of azimuthal angle, $\Psi$, is represented by the solid black points and this has been compared to calculations performed using the FDMNES code \cite{Joly} for the three possible IRs, $\Gamma_{2,6,8}$, described earlier in the text. It is clear that only the $\Gamma_{2}$ representation, associated with an ordered structure where the magnetic moments point along the [001]-axis, can adequately model the data.
\begin{figure*}[htbp]
\centering
\includegraphics[width=1\textwidth,bb=10 330 500 525,clip]{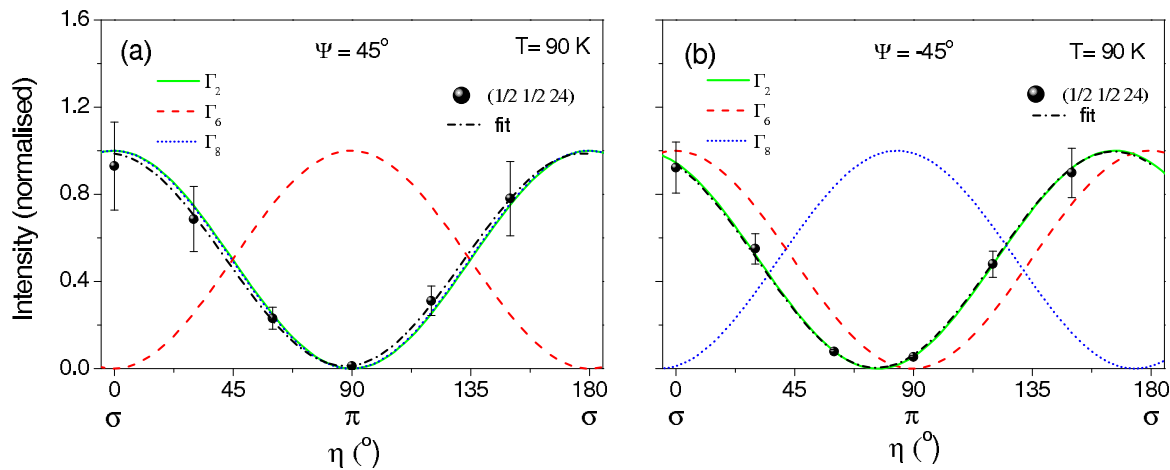}\caption
{\label{figure4}(Color online) Polarisation dependence of the ($\frac{1}{2}$,$\frac{1}{2}$,24) magnetic reflection
for two different azimuthal angles $\Psi$=45$^{\circ}$ (a) and $\Psi$=-45$^{\circ}$ (b) defined with respect to the reference vector (0,1,0). These measurements were made in a horizontal scattering geometry ($\pi$ incoming polarisation) at the Ir L$_3$ edge at a temperature of 90$\,$K. The solid black spheres represent the integrated intensity of the experimental ($\frac{1}{2}$,$\frac{1}{2}$,24) magnetic reflection as a function of the analyser rotation $\eta$. The black dashed line is a fit to equation \ref{equation1} \cite{Hatton}. The green solid line, the red dashed line and the blue dotted line show the FDMNES calculations for the IRs $\Gamma_2$, $\Gamma_6$ and $\Gamma_8$, respectively.}
\end{figure*}

We have also performed a polarisation analysis, using a $\pi$-polarised incident x-ray beam in a horizontal scattering geometry. Figure \ref{figure4} shows calculations for the three IRs, $\Gamma_{2,6,8}$ (green solid, red dashed and blue dotted lines, respectively), and the integrated intensity of the ($\frac{1}{2}$,$\frac{1}{2}$,24) magnetic reflection (black solid points) as a function of the analyser rotation, $\eta$, for two fixed azimuthal angles, 45$^{\circ}$ (a) and -45$^{\circ}$ (b). The dashed black line is a fit to equation \ref{equation1} \cite{Hatton},

\begin{equation}\label{equation1}
I=\frac{I_{0}}{2}|1+\mathrm{P}_{1}\cos(2\eta)+\mathrm{P}_{2}\sin(2\eta)|.
\end{equation}

\noindent P$_{1}$ and P$_{2}$ are Poincar\'{e}-Stokes parameters, where P$_{1}$=(I$_{\sigma}$-I$_{\pi}$)/(I$_{\sigma}$+I$_{\pi}$) and P$_{2}$=(I$_{+45}$-I$_{-45}$)/(I$_{+45}$+I$_{-45}$). For the 45$^{\circ}$ azimuth shown in Figure \ref{figure4} (a), only the calculations for the $\Gamma_{2}$ or $\Gamma_{8}$ IRs well-replicate the experimental data. In (b) (-45$^{\circ}$ azimuth), it is the $\Gamma_{2}$ and $\Gamma_{6}$ IRs that lie close to the dashed black line, although the $\Gamma_{2}$ representation most precisely follows the data. It is clear that only calculations based on a $\Gamma_{2}$ irreducible representation are able to model the experimental data for both the polarisation dependence of the scattered x-rays and the azimuthal dependence. These results indicate that the magnetic moments are aligned along the \textit{c} axis in the low temperature G-type antiferromagnetic phase, at least for the ($\frac{1}{2}$,$\frac{1}{2}$,L=even) domain probed thus far.

\begin{figure}[h!]
\centering
\includegraphics[width=0.75\textwidth,bb=25 230 575 610,clip]{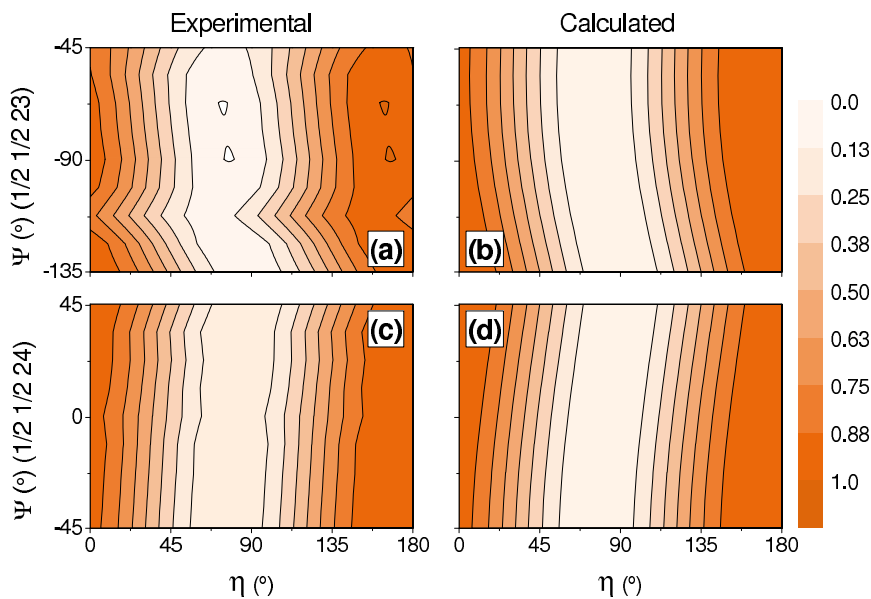}\caption
{\label{figure5}(Color online) The intensity of the($\frac{1}{2}$,$\frac{1}{2}$,23) and the ($\frac{1}{2}$,$\frac{1}{2}$,24) magnetic reflections were fitted to equation \ref{equation1} and are shown in panels (a) and (c), respectively. The experimental data were acquired with a $\pi$-polarised incoming beam at the Ir L$_{3}$ edge at 90$\,$K. Panels (b) and (d) show FDMNES calculations for the ($\frac{1}{2}$,$\frac{1}{2}$,23) magnetic reflection and the ($\frac{1}{2}$,$\frac{1}{2}$,24) magnetic reflection as a function of the analyser rotation $\eta$ and the azimuthal angle $\Psi$.}
\end{figure}

In order to determine the orientation of magnetic moments for both domains, we measured the intensity of ($\frac{1}{2}$,$\frac{1}{2}$,23) and ($\frac{1}{2}$,$\frac{1}{2}$,24) magnetic reflections as a function of the analyser rotation, $\eta$, in a horizontal scattering geometry (with $\pi$ polarised incident x-rays) at T=90$\,$K at the Ir L$_{3}$ edge. Data were collected over a wide range of azimuthal angles, $\Psi$ and fitted to equation (\ref{equation1}), these are shown in panels (a) and (c) of Figure \ref{figure5} for the L=23,24 reflections, respectively. FDMNES calculations based on the $\Gamma_{2}$ representation were performed in order to model this experimental data and these are shown in panels (b) and (d). The data are well-modelled by these calculations, which are based on a two-domain picture of the G-type antiferromagnetic structure with wave vectors $\textbf{k}_{1}$=($\frac{1}{2}$,$\frac{1}{2},$0) and $\textbf{k}_{2}$=($\frac{1}{2}$,-$\frac{1}{2},$0), with moments aligned collinear along the \textit{c} axis.

\begin{figure}[htbp]
\centering
\includegraphics[width=0.75\textwidth,bb=140 300 450 535,clip]{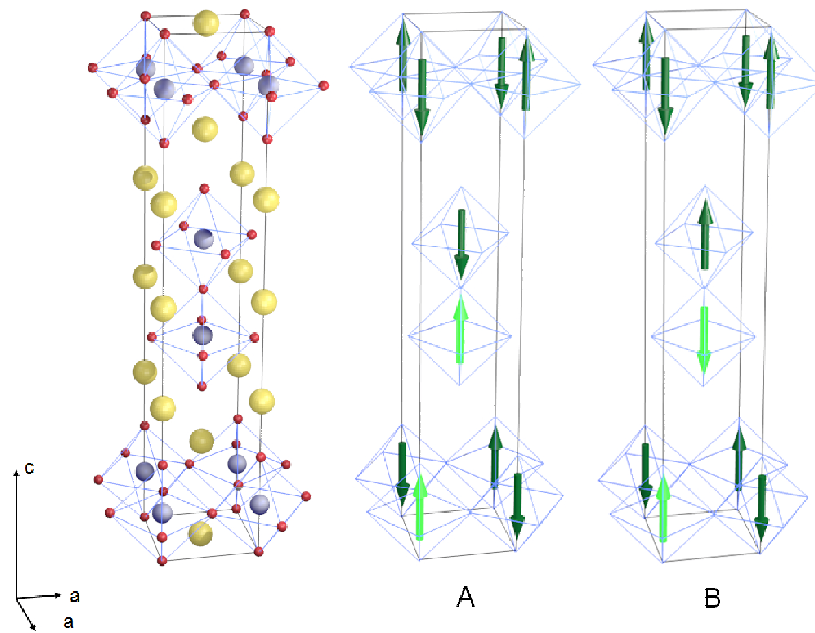}\caption
{\label{figure6}(Color online) The left-hand panel shows a unit cell of the Sr$_{3}$Ir$_{2}$O$_{7}$ crystal structure. Sr atoms (yellow) separate layers of Ir atoms (blue), which reside in the centre of corner sharing oxygen (red) octahedra. The central and right-hand panels sketch the magnetic moments aligned along the \textit{c} axis and situated at the centre of the Ir ions for magnetic structures A and B.}
\end{figure}

Figure \ref{figure6} shows the Sr$_{3}$Ir$_{2}$O$_{7}$ crystal structure (left-hand panel) and representations of the magnetic structures A and B. The magnetic moments are ordered in a G-type antiferromagnetic arrangement with the moments pointing along the \textit{c} axis. Two of the moments have been highlighted in each case to show the difference between the two magnetic wave vectors $\textbf{k}_{1}$ and $\textbf{k}_{2}$, responsible for the domain-like nature of the resonant scattering picture.

Recently, a measurement of the polarisation analysis at two azimuthal angles has been used to infer an orientation of the magnetic moment along the \textit{c} axis, supported by a theoretical discussion including pseudo-dipolar (PD) and Dzyaloshinsky-Moriya interactions in the Hamiltonian \cite{Kim3}. However, this report contains some notable omissions. The magnetic structure presented in Figure 1 of Kim \textit{et al.} \cite{Kim3}, is equivalent to structure B in Figure \ref{figure1} (b) of this article and results in magnetic Bragg peaks at (1,0,L=odd) or (0,1,L=even) in their orthorhombic representation. The orthorhombic orientation is arbitrary of course, for a pseudo-tetragonal system, but having selected an orientation, on exchange of H and K indices, one arrives at a degenerate magnetic structure, precisely that presented in Figure \ref{figure6} (a). There is no mention of the possibility of these two different structures, and in fact there is evidence of intensity for (0,1,L=odd) \textit{and} (0,1,L=even) peaks in the data of Kim \textit{et al.}, shown in Figure 2 (a) of that paper. This is due to the simultaneous illumination of both domains by the footprint of their incident x-ray beam.

This report presents a detailed series of measurements, using a highly focussed incident beam, probing the azimuthal dependence and mapping the polarisation behaviour for domains of magnetic wave vectors $\textbf{k}_{1}$=($\frac{1}{2}$,$\frac{1}{2},$0) and $\textbf{k}_{2}$=($\frac{1}{2}$,-$\frac{1}{2},$0), relating to magnetic structures A and B, respectively. An understanding of this domain behaviour, is particularly important in developing a precise model to describe the magnetic structure. A number of symmetry allowed irreducible representations were determined, using SARA\textit{h} and these were input into FDMNES in order to calculate their azimuthal and polarisation dependence for a given reflection. A comparison of the experimental data with these calculations determined the alignment of the Ir $J_{eff}=\frac{1}{2}$ magnetic moments to be along the \textit{c} axis of a G-type antiferromagnet. This result explains the lack of zero-field magnetic susceptibility, measured along the [001] direction, reported previously \cite{Cao}. However, the origins of the unusual low temperature, ab-plane behaviour, is a more complex problem and remains an intriguing mystery in this compound, the details of which we continue to pursue. This \textit{c} axis alignment of the magnetic moments is different to other layered perovskite iridates \cite{Kim,Okabe} reported to date, and may provide a unique insight into the dominant interactions responsible for the anisotropies in these systems.

\section{Acknowledgements}

We thank the Impact studentship programme, awarded jointly by UCL and Diamond Plc. for funding the thesis work of S. Boseggia. G. Nisbet provided excellent instrument support at the I16 beamline. The research was supported by the EPSRC.

\section{References}

\bibliographystyle{unsrt}
\bibliography{Sr3Ir2O7_rapid}

\begin{thebibliography}{10}

\bibitem{Kim2}
B.~J. Kim, H.~Ohsumi, T.~Komesu, S.~Sakai, T.~Morita, H.~Takagi, and T.~Arima.
\newblock {\em Science}, 323:1329, 2009.

\bibitem{Shitade}
A.~Shitade, H.~Katsura, J.~Kune\v{s}, X-L. Qi, S-C. Zhang, and N.~Hagaosa.
\newblock {\em Phys. Rev. Lett.}, 102:256403, 2009.

\bibitem{Jiang}
H-C. Jiang, Z-C. Gu, X-L. Qi, and S.~Trebst.
\newblock {\em Phys. Rev. B}, 83:245104, 2011.

\bibitem{Machida}
Y.~Machida, S.~Nakatsuji, S.~Onoda, T.~Tayama, and T.~Sakakibara.
\newblock {\em Nature}, 463:210, 2010.

\bibitem{Boseggia}
S.~Boseggia, R.~Springell, H.~C. Walker, A.~T. Boothroyd, D.~Prabhakaran,
  D.~Wermeille, L.~Bouchenoire, S.~P. Collins, and D.~F. McMorrow.
\newblock {\em Phys. Rev. B}, 85:184432, 2012.

\bibitem{Kim3}
J.~W. Kim, Y.~Choi, Jungho Kim, J.~F. Mitchell, G.~Jackeli, M.~Daghofer,
  J.~van~den Brink, G.~Khaliullin, and B.~J. Kim.
\newblock {\em arXiv:1205.4381}, 2012.

\bibitem{Dhital}
C.~Dhital, S.~Khadka, Z.~Yamani, C.~de~la Cruz, T.~C. Hogan S.~M. Disseler,
  M.~Pokharel, K.~C. Lukas, W.~Tan, C.~P. Opeil, Z.~Wang, and S.~D. Wilson.
\newblock {\em arXiv:1206.1006}, 2012.

\bibitem{Jackeli}
G.~Jackeli and G.~Khaliullin.
\newblock {\em Phys. Rev. Lett.}, 102:017205, 2009.

\bibitem{Subramanian}
M.~A. Subramanian, M.~K. Crawford, and R.~L. Harlow.
\newblock {\em Materials Research Bulletin}, 29:645, 1994.

\bibitem{Cao}
G.~Cao, Y.~Xin, C.~S. Alexander, J.~E. Crow, P.~Schlottmann, K.~Crawford, R.~L.
  Harlow, and W.~Marshall.
\newblock {\em Phys. Rev. B}, 66:214412, 2002.

\bibitem{Matsuhata}
H.~Matsuhata, I.~Nagai, Y.~Yoshida, S.~Hara, S.-I. Ikeda, and N.~Shirakawa.
\newblock {\em J. Solid State Chem.}, 177:3776, 2004.

\bibitem{Nagai}
I.~Nagai, Y.~Yoshida, S.-I. Ikeda, H.~Matsuhata, H.~Kito, and M.~Kosaka.
\newblock {\em J. Phys.: Condens. Matter}, 19:136214, 2007.

\bibitem{Wills}
A.~S. Wills.
\newblock Program available from ftp://ftp.ill.fr/pub/dif/sarah/.
\newblock {\em Physica B}, 680:276, 2000.

\bibitem{Joly}
Y.~Joly.
\newblock "x-ray absorption near edge structure calculations beyond the
  muffin-tin approximation".
\newblock {\em Phys. Rev. B}, 63:125120, 2001.

\bibitem{note}
The parameters for this calculation have already been presented in Boseggia
  \textit{et al.} Ref. 5.

\bibitem{Hatton}
P.~D. Hatton, R.~D. Johnson, S.~R. Bland, C.~Mazzoli, T.~A.~W. Beale, C.-H. Du,
  and S.~B. Wilkins.
\newblock {\em J. Magn. Magn. Mater.}, 321:810, 2009.

\bibitem{Kim}
B.~J. Kim, H.~Jin, J.-Y. Kim, B.-G. Park, C.S. Leem, J.~Yu, T.~W. Noh, C.~Kim,
  S.-J. Oh, J.-H. Park, V.~Durairaj, G.~Cao, and E.~Rotenberg.
\newblock {\em Phys. Rev. Lett.}, 101:076402, 2008.

\bibitem{Okabe}
H.~Okabe, M.~Isobe, E.~Takayama-Muromachi, A.~Koda, S.~Takeshita, M.~Hiraishi,
  M.~Miyazaki, R.~Kadono, Y.~Miyake, and J.~Akimitsu.
\newblock {\em Phys. Rev. B}, 83:155118, 2011.

\end{thebibliography}

\end{document}